\documentclass[12pt]{article}
\usepackage{epsfig}
\usepackage{amsmath}

\date{\today}

\newcommand{\be}{\begin{equation}}
\newcommand{\ee}{\end{equation}}
\newcommand{\bea}{\begin{eqnarray}}
\newcommand{\eea}{\end{eqnarray}}
\newcommand{\bml}{\begin{mathletters}}
\newcommand{\eml}{\end{mathletters}}

\begin{document}
\thispagestyle{empty}
\vspace*{1.cm}
\begin{center}
{\Large \bf 
A new spin on black hole hair
}
\\
\vspace{1.5cm}
 {\large Carlos A. R. Herdeiro\footnote{herdeiro@ua.pt} and Eugen Radu\footnote{eugen.radu@ua.pt}} 
\\
\vspace{0.5cm}
\small{
 Departamento de F\'\i sica da Universidade de Aveiro and I3N, 
   \\
   Campus de Santiago, 3810-183 Aveiro, Portugal} 
\end{center}

%\newpage
%\vspace*{1.cm}
\begin{abstract} 
We show that scalar hair can be added to rotating, vacuum black holes of general relativity.
% but only if they are spinning.
%These hairy black holes (HBHs) 
%reduce to spinning smooth scalar field bound states -- \textit{boson stars} -- when the horizon size shrinks to zero. 
These hairy black holes (HBHs) clarify a lingering question concerning gravitational solitons:
  if a black hole can be added at the centre of a boson star, as it typically can for other solitons. 
 We argue that it can, but only if it is spinning. 
 The existence of such  HBHs is related to the Kerr superradiant instability triggered by a massive scalar field. 
This connection leads to the following conjecture: \textit{a (hairless)
   black hole which is afflicted by the superradiant instability of a given 
   field must allow hairy generalizations with that field}.

\end{abstract}

\vspace*{1.cm}

\begin{center}
{\footnotesize{Essay written for the Gravity Research Foundation 2014 Awards for Essays on Gravitation.}}
\\
{\footnotesize{Submitted March 14th, 2014}}

\end{center}
\newpage
\vspace{0.5cm}

%%%%%%%%%%%%%%%%%%%%%%%%%%%%%%%%%%%%%%%%%%%%%%%%%%%%%%%%%%%%%%%%%%%%%%%%%%%%%%
%%%%%%                    motivation
%%%%%%%%%%%%%%%%%%%%%%%%%%%%%%%%%%%%%%%%%%%%%%%%%%%%%%%%%%%%%%%%%%%%%%%%%%%%%%

%%%%%%%%%%%%%%%%%%%%%%%%%
\section{Prologue: black hole baldness}
%%%%%%%%%%%%%%%%%%%%%%%%%

\textit{``In my entire scientific life,
extending over forty-five years, the most shattering
experience has been the realization that an exact solution
of Einstein's equations of general relativity, discovered
by the New Zealand mathematician, Roy Kerr, provides the
absolutely exact representation of untold numbers of
massive black holes that populate the universe.''}  

\bigskip

This quote, by S. Chandrasekhar~\cite{Chandra}, highlights a central result from black hole (BH) theory: the uniqueness theorems of vacuum Einstein's gravity~\cite{Robinson:2004zz}. Such results endow the BH concept with such an extraordinary elegance and simplicity that Chandrasekhar could not help feeling a sense of awe. The same idea became carved in stone by John Wheeler's statement  {\it ``BHs have no hair''}~ \cite{Wheeler}.  In a nutshell, that whatever matter originates the BH, all its information -- to which `hair' provides an image -- disappears, except for a small set of asymptotically measurable quantities.  

In this essay, we will revisit the `BH no-hair' idea and show that, albeit compelling, there is new evidence to reconsider it.

\newpage

%%%%%%%%%%%%%%%%%%%%%%%%%
\section{Hairy black holes and horizonless solitons}
%%%%%%%%%%%%%%%%%%%%%%%%%
 At present,  ``hair'' is used in BH physics as a colloquial term to describe any measure -- 
 beyond those subjected to a Gauss law, such as mass, angular momentum and electric/magnetic charges -- 
 needed to fully describe the BH. 
 As such, there are by now a number of counterexamples to the no-hair idea, 
for various nonlinear matter sources coupled to general relativity
in a four dimensional, asymptotically flat spacetime 
(for reviews, see 
\cite{Bizon:1994dh,Bekenstein:1996pn,Volkov:1998cc}). 
These counterexamples are typically unstable and/or occur 
 in rather exotic theories, but they illustrate the $mathematical$ limitations of the no-hair idea.
 
An analysis of the known hairy black holes (HBHs) shows that these solutions, in a given model, typically occur together with horizonless soliton-like configurations, obtained from the HBH in the limit of vanishing horizon size. Conversely, a rule of thumb is that  if some type of matter allows for solitons when coupled to gravity, then it also allows for HBHs. 
This state of affairs has led to a description of some HBHs
as  bound states of 
ordinary BHs (without hair) and solitons, 
within the isolated horizon formalism 
\cite{Ashtekar:2000nx}. 

From the above, it appears rather mysterious that 
there are no known asymptotically flat BHs with scalar hair,
since a massive complex scalar field  can condense to
 form smooth horizonless bound states: {\it boson stars} (BSs).
BSs exist due to a balance between their self-generated gravity and the dispersive effect 
due to the wave character of the scalar field  \cite{Kaup:1968zz}.
% which must have a harmonic time dependence 
 Such solitons are argueably the physically most interesting gravitating solitons, 
  considered as possible BH mimickers and dark matter candidates 
(see $e.g.$ the recent  review \cite{Liebling:2012fv}).

The original no-hair theorems do not cover scalar fields with a harmonic time dependence,
 but the  results in \cite{Pena:1997cy} prove the absence of 
BH generalizations of {\it spherically symmetric} BSs. 
 This led to a widespread believe that it is not possible to add an horizon in the interior of 
\textit{any} BS,
 without trivializing the scalar field.
  This believe, however, turns out to be incorrect. 
%As we show next, 
%the soliton solutions (\textit{i.e.} boson stars) of the Einstein--Klein-Gordon (EKG) theory 
%allow for HBHs generalizations, in close analogy with other models.

 %%%%%%%%%%%%%%%%%%%%%%%%%%
\section{\textit{Spinning} boson stars may wear black...}
%%%%%%%%%%%%%%%%%%%%%%%%%%
 The crucial new ingredient to obtain HBHs with scalar hair is \textit{spin}. HBHs must be studied using numerical methods, since no exact analytic solution is known for  BSs, even with spherical symmetry.
We found HBH solutions by solving numerically the field equations with a metric ansatz
%with suitable boundary conditions.
%These HBHs are axially symmetric, 
%being described by a line element 
with two Killing vectors:
$\xi=\partial_t$ 
and 
$\eta=\partial_\varphi$. These are not, however, Killing vectors of the full solution;
  the scalar field $\Psi$ depends on both $\varphi$ and $t$ through a phase. Thus HBHs have a single Killing vector field, c.f.~\cite{Dias:2011at}. The full ansatz reads: 
  %\footnote{Note that the Kerr metric
%can also be written in this form.}:
%
\begin{eqnarray}
\label{metric-ansatz}
&ds^2=e^{2F_1\left(r,\theta\right)}\left(\frac{dr^2}{1-\frac{r_H}{r} }
+r^2 d\theta^2\right)+e^{2F_2(r,\theta)}r^2 \sin^2\theta (d\varphi-W(r,\theta) dt)^2-e^{2F_0(r,\theta)} \left(1-\frac{r_H}{r}\right) dt^2 ,
\nonumber
\\
\label{scalar-ansatz}
&\Psi=\phi(r,\theta)e^{i(m\varphi-w t)}.
%\nonumber
%~{\rm with}~~~~N=1-\frac{r_H}{r}.~~~~~{~~~~~~~~~~~~~}
\end{eqnarray}  
$w>0$ is the frequency and $m=\pm 1,\pm 2$\dots
is the azimuthal winding number. 
As for BSs, the existence of HBHs requires
the scalar field to be massive. $\Psi$ may also possess a self-interacting potential; but this is not mandatory.
$r_H\geq 0$ fixes the position of
the event horizon; thus the solitonic limit of HBHs is obtained by taking $r_H\rightarrow 0$ and yields the spinning
BSs in \cite{s1,Yoshida:1997qf}. The near horizon expansion of the solutions imposes that the horizon angular velocity obeys
 \begin{eqnarray}
\label{cond}
\Omega_H=\frac{w}{m}.
\end{eqnarray}
This implies that there is no flux of scalar field
into the HBH, $\chi^{\mu}\partial_\mu \Psi=0$, where $\chi =\xi+\Omega_H \eta$ is the null horizon generator. 
%In the static limit, this condition implies immedi
%The scalar field, however, does not vanish on the horizon.

%With the ansatz (\ref{metric-ansatz}),  
%the  EKG equations reduce to five second order  coupled non-linear 
%elliptic partial differential equations,  which we have solved
%subject to suitable boundary conditions [REF]. 

%, $\phi(r_H,\theta)\neq 0$.
%Also, the Killing vector $\chi =\xi+\Omega_H \eta$ is orthogonal and null on the horizon.

The HBH solutions were obtained by slowly increasing $r_H$ from zero, for given $m,w$~\cite{HR}. 
%By increasing $r_H$ from zero, we obtain a set of  HBHs solutions with $\Omega_H$
%fixed by (\ref{cond}). 
They  possess  a nonvanishing 
scalar field on and outside a (regular) horizon,
providing perhaps the simplest violation of the no-hair idea.
%The HBHs with a given event horizon coordinate radius $r_H$
%exist for a limited range of frequencies. 
The scalar field surfaces of constant
energy density have a toroidal topology near the horizon and a spherical topology asymptotically  -- Fig. 1.

%%%%%%%%%%%%%%%%%%%%%%%%%%%%%%%%%%%%%%%%%%%%%%%%%%%%%%%%%%%%%%%%%%%%%%%
\vspace{1.cm} 
 \setlength{\unitlength}{1cm}
\begin{picture}(8,6)
\put(-0.5,0.5){\epsfig{file=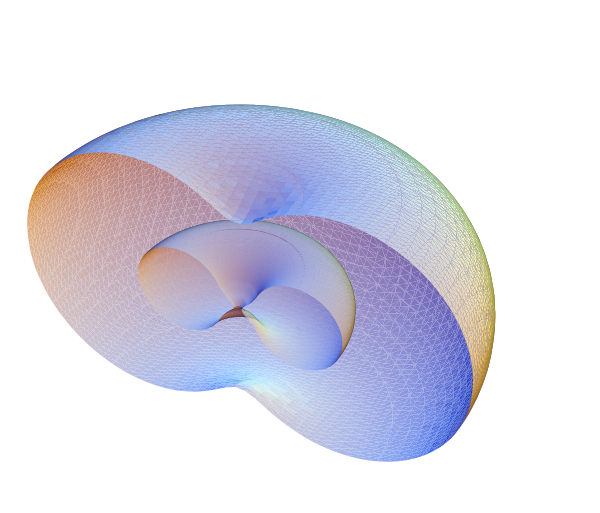,width=7cm}}
\put(8,0.0){\epsfig{file=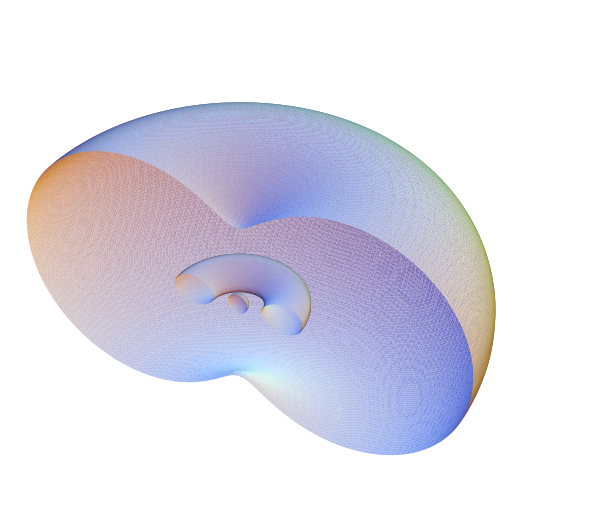,width=8cm}}
\end{picture}
 \\
{\small {\bf Figure 1.}
{\small  
Isosurfaces of constant energy density,
%$\rho=-T^t_t$ with 
$E_1=0.17$ (outside shell) and $E_2=3$, for a BS (left)
and a HBH (right), with parameters $m=1$, $w/\mu=0.81$ and 
$(M\mu,J\mu^2,Q\mu^2)$= 
$( 0.65,0.42,0.42)$ and $(0.72,0.49,0.46)$, respectively. 
$\mu$ is the scalar field mass and Newton's constant is set to $G=1$.
 The event horizon is the sphere at the center of the right plot. 
 }
%%%%%%%%%%%%%%%%%%%%%%%%%%%%%%%%%%%%%%%%%%%%%%%%%%%%%%%%%%%%%%%%%%%%%%%

%\newpage

 %%%%%%%%%%%%%%%%%%%%%%%%%%
\section{... \textit{i.e.} \textit{spinning} black holes may wear scalar hair}
%%%%%%%%%%%%%%%%%%%%%%%%%%
Since BHs can be added at the center 
of spinning BSs: 1) \textit{how can we fully describe these solutions?} and  2) \textit{do these solutions connect continuously with (hairless) Kerr BHs?}

As for the first question, the only global charges of a HBH, computed using a Gauss law, are the mass $M$ and angular momentum $J$. In contrast to the Kerr case, they fail to fully specify the HBH solution. In fact, HBHs can even coexist with Kerr BHs in some region of the $(M,J)$ plane, providing a new example of non-uniqueness. Thus, to specify the solution, we add an extra quantity: the global Noether charge
\begin{eqnarray}
\label{Q}
Q=\int_{\Sigma}dr d\theta d\varphi ~j^t \sqrt{-g}, \qquad {\rm where} \ \ j^a=-i (\Psi^* \partial^a \Psi-\Psi \partial^a \Psi^*),
\end{eqnarray}
($j^a$ is a conserved current), which provides a quantitative measure of the scalar field outside the horizon.
% but cannot be reduced to some scalar flux at infinity. 
Our results indicate the $(M,J,Q)$ specify a unique solution, after fixing $m,w$ and the number of scalar field nodes. 
%(note that  $J/Q=m$ in the soliton limit \cite{s1,Yoshida:1997qf}). 

The answer to the second question reveals another feature
which may have far-reaching implications.
For any given $m$, 
the HBHs are indeed connected to a sub-family of Kerr BHs with a particular relation $M=M(J)$~\cite{HR}. It is straightforward to specify a physical requirement for this sub-family. Approaching the Kerr limit, 
 the scalar field becomes arbitrarily small and the geometry arbitrarily close to Kerr. The hair can therefore be seen as scalar \textit{bound states} -- since they have a time independent energy density and an exponential spatial decay -- of the Klein-Gordon equation in the Kerr background. Thus, the Kerr sub-family connected to HBHs must support scalar bound states with the given $m$. 
 
Schwarzschild BHs do not support scalar bound state solutions of the Klein Gordon equation; they only allow  modes that are slowly decaying into the BH. Kerr BHs, on the other hand, support both decaying but also -- because of the \textit{superradiant instability}~ \cite{Press:1972zz} -- growing modes.  Bound states exist at the threshold of the instability. They obey~\eqref{cond} and define, for each $m$, precisely the aforementioned Kerr sub-family.

Thus, the scalar hair is the non-linear realization of the bound states obtained in linear theory (see~\cite{Hod:2012px} for a discussion of bound states for extremal Kerr). And HBHs branch off from the Kerr family at the threshold of the superradiant instability.
\section{A recipe to grow hair}
%%%%%%%%%%%%%%%%%%%%%%%%%%% 

The occurance of a new branch of solutions at the threshold of a classical instability is a well-known
feature of BH physics. The Gregory-Laflamme instability~\cite{Gregory:1993vy} is a classical example.
The previous section established the same pattern for the superradiant instability.  
Thus, more than an example of HBH solutions, the discussion presented in this essay provides a \textit{mechanism} for growing hair on BHs:

\bigskip

\textit{A (hairless) BH which is afflicted by the superradiant instability of a given field must allow hairy generalizations with that field.} 

\bigskip

%The existence of BHs with scalar hair in asymptotically AdS spacetimes also supports this idea~\cite{Dias:2011at}. 
 
%In fact there are other examples in the literature that support this mechanism. Whereas for asymptotically flat solutions,  the black hole spin and the field's mass are mandatory these requirements can be circumvented in Anti-de Sitter space (AdS). There, the scalar field mass can be replaced by AdS asymptotics to yield the necessary confinment for the existence of bound states [??]. Moreover, the black hole spin can be replaced by charge. Charged black holes in AdS are afflicted by a superradiant instability triggered by charged scalar fields. Consequently hairy black holes exist. These HBHs have played an important role in the AdS-Condensed Matter duality. 

% \newpage
 
 %%%%%%%%%%%%%%%%%%%%%%%%%%%  
\section{Epilogue: how simple is the Universe?}
%%%%%%%%%%%%%%%%%%%%%%%%%%%
 
Is Chandrasekhar's realization quoted at the beginning of this essay at risk? 
In other words, in view of these new families of HBHs is it possible 
that the BHs populating the Universe will be different from the Kerr solution? The answer is intimately related to two other questions. Firstly: are there any fields in nature that can excite a superradiant instability of Kerr BHs? And secondly: do the HBHs presented here -- or their generalizations associated to other fields -- play a role, as transient or final states in the \textit{dynamical} development of that superradiant instability?
 
 Understanding these questions will clarify if Chandrasekhar's amazing realization is justified or if Nature is not that simple.

\newpage

%%%%%%%%%%%%%%%%%%%%%%%%%%%%%%%%%%%%%%%%%%%%%%%%%%%%%%%%%%%%%%%%%%%%%%%%%%%%%%

%%%%%%%%%%%%%%%%%%%%%%%%%%%%%%%%%%%%%%%%%%%%%%%%%%%%%%%%%%%%%%%%%%%%%%%%%%%%%%

\end{document}